\documentclass{article}
\usepackage{spconf,amsmath,graphicx, hyperref, amssymb, latexsym}
\usepackage{cite, algorithmic, footmisc, mathtools}
\usepackage{subcaption}
\usepackage{pgfplots}
\pgfplotsset{compat=1.17}
\usepgfplotslibrary{groupplots,dateplot}
\usepackage{balance}

\usepackage[pscoord]{eso-pic}
\newcommand{\placetextbox}[3]{
\setbox0=\hbox{#3}
\AddToShipoutPictureFG*{
\put(\LenToUnit{#1\paperwidth},\LenToUnit{#2\paperheight}){\vtop{{\null}\makebox[0pt][c]{#3}}}}
}

\title{Color Learning for Image Compression}
\name{Srivatsa Prativadibhayankaram, Thomas Richter,  Heiko Sparenberg, Siegfried Foessel}
\address{\textit{Moving Picture Technologies}, Fraunhofer Institute for Integrated Circuits IIS, 
	\\ Erlangen, Germany}

\begin{document}

\maketitle

\begin{abstract}
Deep learning based image compression has gained a lot of momentum in recent times. To enable a method that is suitable for image compression and subsequently extended to video compression, we propose a novel deep learning model architecture, where  the task of image compression is divided into two sub-tasks, learning structural information from luminance channel and color from chrominance channels. The model has two separate branches to process the luminance and chrominance components. The color difference metric CIEDE2000 is employed in the loss function to optimize the model for color fidelity. We demonstrate the benefits of our approach and compare the performance to other codecs. Additionally, the visualization and analysis of latent channel impulse response is performed.
\end{abstract}

\placetextbox{0.5}{0.08}
{\fbox{\parbox{\dimexpr\textwidth-2\fboxsep-2\fboxrule\relax}{\footnotesize \centering \copyright 2023 IEEE. Personal use of this material is permitted. Permission from IEEE must be obtained for all other uses, in any current or future media, including reprinting/republishing this material for advertising or promotional purposes, creating new collective works, for resale or redistribution to servers or lists, or reuse of any copyrighted component of this work in other works. \textbf{Accepted paper - ICIP 2023}}}}

\vspace{-8pt}
\section{Introduction}
\label{sec:intro}
\vspace{-4pt}
Image compression is a high impact technology that minimizes resources for transmission bandwidth and storage. A conventional image codec such as JPEG\cite{125072} uses a block based transform coding approach. The images are partitioned into blocks and transformed into the frequency domain using the discrete cosine transform (DCT), which results in energy compaction. The transformed image is then quantized and entropy coded into a compact binary representation. Video codecs such as HEVC\cite{6316136} and VVC\cite{9503377} contain additional complex blocks for motion prediction and compensation.

The optimization of an image codec is aimed at balancing the trade-off between rate and distortion, termed rate-distortion optimization (RDO). A suitable encoding that results in the least cost is chosen and the corresponding parameters are shared with the decoder, in order to perform reconstruction with a high fidelity. It can be formulated as minimizing loss $L$,
\vspace{-4pt}
\begin{equation}
	\begin{aligned}
		\mathrm{min}\{L\} \text{ with } L = R + \lambda \cdot D,
		\label{eqn:rdo}
	\end{aligned}
\end{equation}
where the rate $R$ is measured in bits per pixel (bpp) and the distortion $D$ is computed by an objective quality metric, and $\lambda$ is the Lagrangian multiplier. Some of the most commonly used metrics for distortion are mean squared error (MSE), peak signal to noise ratio (PSNR), and multi scale structural similarity index (MS-SSIM)\cite{wang_multiscale_2003}. 

Using the same principle of redundancy reduction and representing images in a compact form, there are many deep neural network models that employ non-linear transform coding\cite{balleend}, where the neural networks are trained end-to-end with the encoder performing non-linear dimensionality reduction of the input image into a latent representation and the decoder transforming the latent back into image space. Recently many learning based image codecs have been presented \cite{Gao2021c, cheng_learned_2020, Brand2022a}. The model in \cite{Ma2022a} being on par with the intra-coding mode of VVC. In most of these codecs, we find an architecture that is based on variational autoencoders (VAE), which has an encoder-decoder structure that fits well to the task of compression. While the models are optimized with RDO, we also observe that perceptual metrics based on deep features such as LPIPS \cite{zhang_unreasonable_2018} and adversarial loss are also used as part of loss function. Although most learned image codecs operate in the RGB color space, recent methods in \cite{10018070} and \cite{10017994} use YUV or YCbCr.

Our contributions in this work are as follows. We propose a deep neural network model that makes use of YUV color space in order to exploit the luminance channel for structural information and chrominance channels for color information. In order to optimize the model for color fidelity, we use an explicit metric for color difference - CIEDE2000 \cite{sharma2005ciede2000} in the RDO based loss function.  Moreover, to study the effect of separating structure and color information, the channel impulse responses of the proposed model are analyzed for the luminance and chrominance components by extending the method described in \cite{10018031}.
\vspace{-8pt}
\section{Deep Neural Network based Color Learning}
\label{sec:core_cl}
\vspace{-4pt}
\begin{figure*} 
	\centering
	\resizebox{!}{0.32\textwidth}{\includegraphics{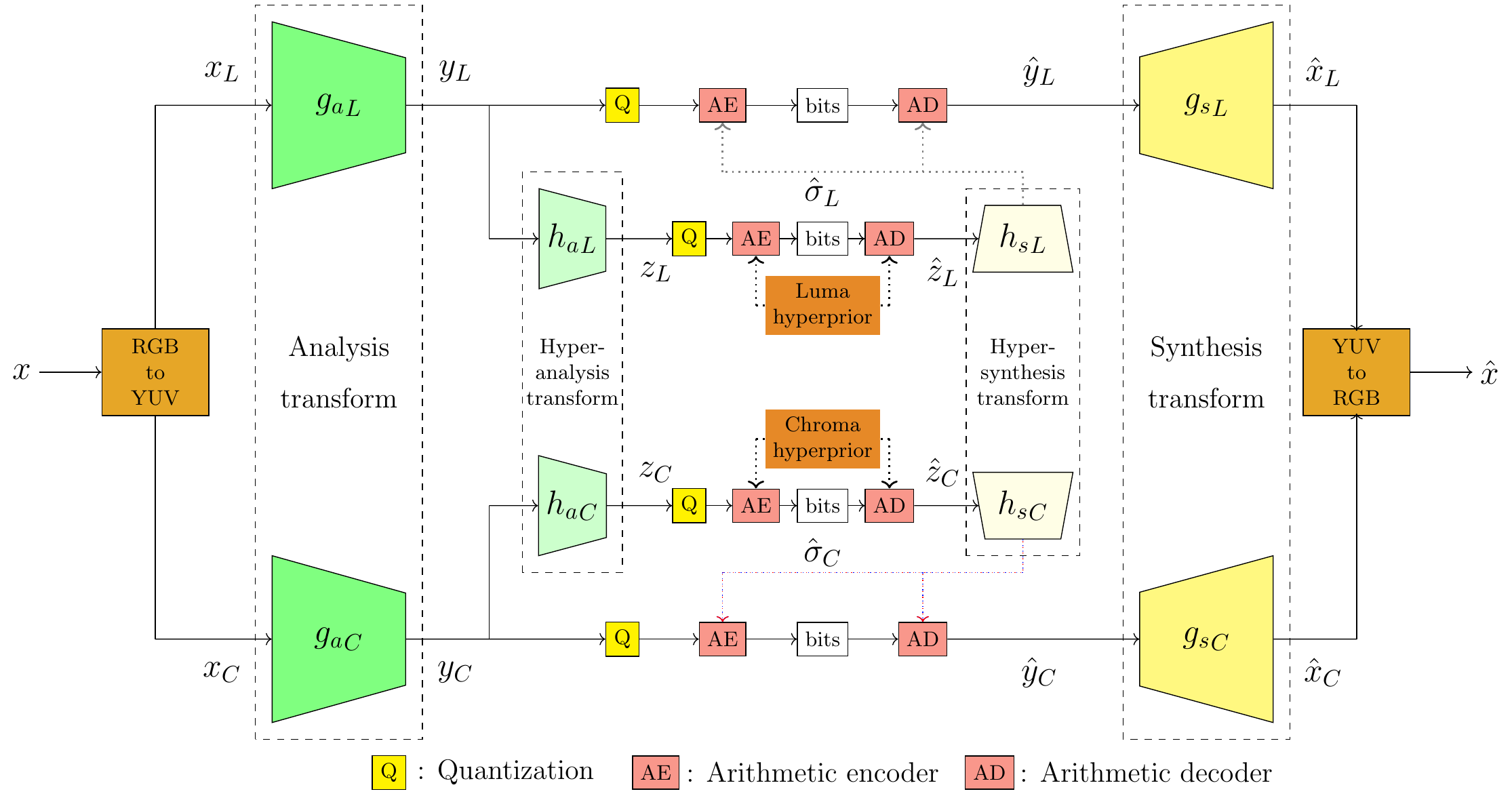}}
	\caption{{Network architecture of the proposed color learning model.} The upper part indicates the luminance branch and the lower part is the chrominance branch.}\label{fig:model}
	\vspace{-16pt}
\end{figure*}
The human visual system is more sensitive to brightness changes in comparison to colors \cite{wandell1995foundations}. Taking this into consideration, the proposed deep learning model processes brightness (luminance) and color (chrominance) components separately for the case of learning based end-to-end image compression. We use the term color learning to illustrate the idea of enabling our model to have color fidelity by employing a metric in the form of CIEDE20000 \cite{sharma2005ciede2000}. In this section we describe the model and its optimization, followed by the channel impulse response computation process and finally the implementation details.

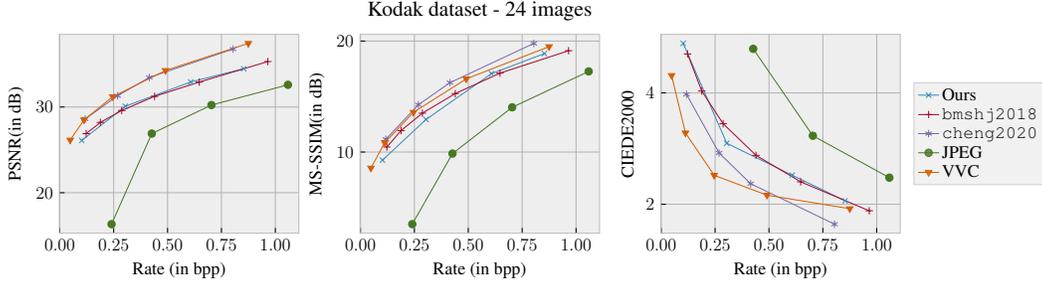
\begin{figure*}[!h]
	\centering 
	\resizebox{0.78\textwidth}{!}{
\begin{tikzpicture}
	
	\definecolor{chocolate213940}{RGB}{213,94,0}
	\definecolor{darkgray178}{RGB}{178,178,178}
	\definecolor{darkolivegreen7012033}{RGB}{70,120,33}
	\definecolor{firebrick166640}{RGB}{166,6,40}
	\definecolor{gainsboro}{RGB}{220,220,220}
	\definecolor{lightgray204}{RGB}{204,204,204}
	\definecolor{silver188}{RGB}{188,188,188}
	\definecolor{slategray122104166}{RGB}{122,104,166}
	\definecolor{steelblue52138189}{RGB}{52,138,189}
	\definecolor{whitesmoke238}{RGB}{238,238,238}
	\Large
	\begin{groupplot}[group style={group size=3 by 1, horizontal sep=1.75cm}]
		\nextgroupplot[
		axis background/.style={fill=whitesmoke238},
		axis line style={silver188},
		tick pos=left,
		x grid style={darkgray178},
		xlabel={Rate (in bpp)},
		xmajorgrids,
		xmin=-0.0022690245, xmax=1.1090373145,
		xtick style={color=black},
		xtick={0,0.25,0.5,0.75,1},
		xticklabels={
			\(\displaystyle {0.00}\),
			\(\displaystyle {0.25}\),
			\(\displaystyle {0.50}\),
			\(\displaystyle {0.75}\),
			\(\displaystyle {1.00}\)
		},
		y grid style={darkgray178},
		ylabel={PSNR(in dB)},
		ymajorgrids,
		ymin=15.276710555, ymax=38.474225665,
		ytick style={color=black},
		ytick={10,20,30,40},
		yticklabels={
			\(\displaystyle {10}\),
			\(\displaystyle {20}\),
			\(\displaystyle {30}\),
			\(\displaystyle {40}\)
		}
		]
		\addplot [thick, steelblue52138189, mark=x, mark size=3, mark options={solid}]
		table {%
			0.101962619 26.09830171
			0.304053413 30.08477592
			0.607754178 32.91841704
			0.853075663 34.46104171
		};
		\addplot [thick, firebrick166640, mark=+, mark size=3, mark options={solid}]
		table {%
			0.122387455776334 26.8910002708435
			0.188215353836616 28.19975233078
			0.287325574085116 29.5900418758392
			0.439705594132344 31.2396558920542
			0.647166167696317 32.9057821432749
			0.965321307380994 35.3013919989268
		};
		\addplot [thick, slategray122104166, mark=asterisk, mark size=3, mark options={solid}]
		table {%
			0.117429048 28.60005073
			0.268648766 31.31135304
			0.415263219 33.44686715
			0.804055966 36.76113274
		};
		\addplot [thick, darkolivegreen7012033, mark=*, mark size=3, mark options={solid}]
		table {%
			0.240111457 16.33114306
			0.426833259 26.89888
			0.703613281 30.22622045
			1.05852339 32.585754
		};
		\addplot [thick, chocolate213940, mark=triangle*, mark size=3, mark options={solid,rotate=180}]
		table {%
			0.0482449 26.14484539
			0.112495422 28.49302175
			0.245816549 31.19987372
			0.490545485 34.26153257
			0.874810113 37.41979316
		};
		
		\nextgroupplot[
		axis background/.style={fill=whitesmoke238},
		axis line style={silver188},
		tick pos=left,
		title={\Large {\LARGE Kodak dataset - 24 images}},
		x grid style={darkgray178},
		xlabel={Rate (in bpp)},
		xmajorgrids,
		xmin=-0.0022690245, xmax=1.1090373145,
		xtick style={color=black},
		xtick={0,0.25,0.5,0.75,1},
		xticklabels={
			\(\displaystyle {0.00}\),
			\(\displaystyle {0.25}\),
			\(\displaystyle {0.50}\),
			\(\displaystyle {0.75}\),
			\(\displaystyle {1.00}\)
		},
		y grid style={darkgray178},
		ylabel={MS-SSIM(in dB)},
		ymajorgrids,
		ymin=2.68667780010524, ymax=20.6041800454937,
		ytick style={color=black},
		ytick={0,10,20,30},
		yticklabels={
			\(\displaystyle {0}\),
			\(\displaystyle {10}\),
			\(\displaystyle {20}\),
			\(\displaystyle {30}\)
		}
		]
		\addplot [thick, steelblue52138189, mark=x, mark size=3, mark options={solid}]
		table {%
			0.101962619 9.27087806598775
			0.304053413 12.9344306438835
			0.607754178 17.0634911644116
			0.853075663 18.8736500771384
		};
		\addplot [thick, firebrick166640, mark=+, mark size=3, mark options={solid}]
		table {%
			0.122387455776334 10.452419176473
			0.188215353836616 11.9345906639915
			0.287325574085116 13.504577936533
			0.439705594132344 15.2668401336408
			0.647166167696317 17.1097146449593
			0.965321307380994 19.1301079070258
		};
		\addplot [thick, slategray122104166, mark=asterisk, mark size=3, mark options={solid}]
		table {%
			0.117429048 11.1542097084031
			0.268648766 14.2656720848146
			0.415263219 16.2514124361455
			0.804055966 19.7897481252488
		};
		\addplot [thick, darkolivegreen7012033, mark=*, mark size=3, mark options={solid}]
		table {%
			0.240111457 3.50110972035017
			0.426833259 9.86232379601698
			0.703613281 14.0303979911009
			1.05852339 17.2636889039262
		};
		\addplot [thick, chocolate213940, mark=triangle*, mark size=3, mark options={solid,rotate=180}]
		table {%
			0.0482449 8.55762649098114
			0.112495422 10.8455232115259
			0.245816549 13.5761193814512
			0.490545485 16.5912563121794
			0.874810113 19.4988314300131
		};
		
		\nextgroupplot[
		axis background/.style={fill=whitesmoke238},
		axis line style={silver188},
		legend cell align={left},
		legend style={
			fill opacity=0.8,
			draw opacity=1,
			text opacity=1,
			at={(1.6,0.5)},
			anchor=east,
			draw=lightgray204,
			fill=whitesmoke238
		},
		tick pos=left,
		unbounded coords=jump,
		x grid style={darkgray178},
		xlabel={Rate (in bpp)},
		xmajorgrids,
		xmin=-0.0022690245, xmax=1.1090373145,
		xtick style={color=black},
		xtick={0,0.25,0.5,0.75,1},
		xticklabels={
			\(\displaystyle {0.00}\),
			\(\displaystyle {0.25}\),
			\(\displaystyle {0.50}\),
			\(\displaystyle {0.75}\),
			\(\displaystyle {1.00}\)
		},
		y grid style={darkgray178},
		ylabel={CIEDE2000},
		ymajorgrids,
		ymin=1.47856880825, ymax=5.05274723475,
		ytick style={color=black},
		ytick={0,2,4,6},
		yticklabels={
			\(\displaystyle {0}\),
			\(\displaystyle {2}\),
			\(\displaystyle {4}\),
			\(\displaystyle {6}\)
		}
		]
		\addplot [thick, steelblue52138189, mark=x, mark size=3, mark options={solid}]
		table {%
			0.101962619 4.890284579
			0.304053413 3.095231221
			0.607754178 2.521409721
			0.853075663 2.060963129
		};
		\addlegendentry{Ours}
		\addplot [thick, firebrick166640, mark=+, mark size=3, mark options={solid}]
		table {%
			0.122387455776334 4.70026633143425
			0.188215353836616 4.03393138448397
			0.287325574085116 3.44657446444035
			0.439705594132344 2.8787375887235
			0.647166167696317 2.40204610923926
			0.965321307380994 1.88158416748047
		};
		\addlegendentry{\texttt{bmshj2018}}
		\addplot [thick, slategray122104166, mark=asterisk, mark size=3, mark options={solid}]
		table {%
			0.117429048 3.97457995
			0.268648766 2.920498694
			0.415263219 2.367824644
			0.804055966 1.641031464
		};
		\addlegendentry{\texttt{cheng2020}}
		\addplot [thick, darkolivegreen7012033, mark=*, mark size=3, mark options={solid}]
		table {%
			0.240111457 nan
			0.426833259 4.792341759
			0.703613281 3.230134184
			1.05852339 2.477046217
		};
		\addlegendentry{JPEG}
		\addplot [thick, chocolate213940, mark=triangle*, mark size=3, mark options={solid,rotate=180}]
		table {%
			0.0482449 4.313641921
			0.112495422 3.281450525
			0.245816549 2.519429068
			0.490545485 2.162770063
			0.874810113 1.919252791
		};
		\addlegendentry{VVC}
	\end{groupplot}
	
\end{tikzpicture}}
	\caption{R-D curves of the proposed model (ours), \texttt{bmshj2018}\cite{balle2018variational}, \texttt{cheng2020}  \cite{cheng_learned_2020}, JPEG \cite{125072} and VVC \cite{9503377} for the \emph{Kodak} dataset.}
	\label{fig:rd_curves}
		\vspace{-16pt}
\end{figure*}

\vspace{-8pt}
\subsection{Model and workflow}\label{ssec:model}
\vspace{-4pt}
The proposed model consists of the following blocks, in the order of execution - RGB to YUV conversion, analysis transform, hyperencoder, hyperdecoder, entropy coding, synthesis transform and YUV to RGB conversion block. It is based on the end-to-end model in \cite{balle2018variational}. The network architecture of the model is shown in Fig. \ref{fig:model}. For chroma-subsampled variants of YUV, suitable downsampling and upsampling blocks can be added. The encoder and the decoder have two branches. The upper branch is for the luminance (Y) and the lower branch for chrominance (UV). The hyperencoders and hyperdecoders assist entropy coding by estimating the distribution of the latents that is learnt during training by means of hyperprior.\\ \\
\textbf{Model: }The analysis transforms for luminance and chrominance are represented by ${g_{a}}_{L}(\cdot)$ and ${g_{a}}_{C}(\cdot)$ respectively. The hyper analysis transforms are indicated by ${h_{a}}_{L}(\cdot)$ and ${h_{a}}_{C}(\cdot)$, which are complemented by the hyper synthesis transforms ${h_{s}}_{L}(\cdot)$ and ${h_{s}}_{C}(\cdot)$. Finally, ${g_{s}}_{L}(\cdot)$ and ${g_{s}}_{C}(\cdot)$ indicate the synthesis transforms. We mainly use convolutional layers with generalized divisive normalization (GDN) \cite{balle2016density} for non-linearity. We also include attention mechanism in the form of convolutional block attention module (CBAM) \cite{woo2018cbam} in the analysis and synthesis transforms for better feature extraction. The analysis transforms and the synthesis transforms have an overall upsampling and downsampling factor of 16. To assign higher importance to luminance, the upper or the luminance branch is designed to have 128 channels. The chrominance or lower branch has 64 channels. The hyperpriors are also initialized accordingly for luminance and chrominance channels as noisy normal distributions. The main entropy bottlenecks are considered to be zero-mean Gaussian distributions whose standard deviations are predicted by the hypersynthesis transforms, as described in \cite{balle2018variational}. \\
\textbf{Encoding: } An RGB image $x$ is first converted into YUV, split into $x_{L}$(luminance), that has a single channel and $x_{C}$ (chrominance) consisting of two channels. These form the inputs to the analysis transforms ${g_{a}}_{L}(\cdot)$ and ${g_{a}}_{C}(\cdot)$, which transform them to latents $y_{L}$ and $y_{C}$ respectively. The hyperanalysis transforms ${h_{a}}_{L}(\cdot)$ and ${h_{a}}_{C}(\cdot)$ convert latents into hyperlatents $z_{L}$ and $z_{C}$. These are quantized and entropy coded using hyperpriors. The quantized  hyperlatents $\hat{z}_{L}$ and $\hat{z}_{C}$,  are used by the hypersyntheis transforms ${h_{s}}_{L}(\cdot)$ and ${h_{s}}_{C}(\cdot)$ to estimate the latent distributions $\hat{\sigma}_{L}$ and $\hat{\sigma}_{C}$ for luminance and chrominance respectively. The estimated distributions are used to encode the latents as bitstreams. Effectively, the bitstream consists of four components - quantized and entropy coded latents of luminance ($y_{L}, z_{L}$) and chrominance ($y_{C}, z_{C}$). The decoding is the inverse process that finally results in the reconstructed image $\hat{x}$.

\vspace{-8pt}
\subsection{Optimization}
\vspace{-4pt}
The resulting bitrate from a model can be attributed to the choice of the Lagrangian multiplier $\lambda$. 

To estimate distortion, the reconstructed image $\hat{x}$ is compared to the original image $x$ through a suitable metric. Let us denote the effective model parameters of the encoder by $\boldsymbol{\theta}$ and that of the decoder by $\boldsymbol{\phi}$. In addition to MSE and MS-SSIM, we use a color difference metric CIEDE2000, denoted by $\Delta E_{00}^{12}$ for color fidelity. This metric employs three components - luminance, chrominance and hue for measuring the color difference between two values. The detailed derivation of the metric can be found in \cite{sharma2005ciede2000}. A lower value of CIEDE2000 indicates a lower color difference. The loss function for the proposed model can be formulated by extending Equation \ref{eqn:rdo} as,
\vspace{-2pt}
\begin{equation}
	\begin{aligned}
		\mathrm{min}_{\boldsymbol{\theta}, \boldsymbol{\phi}}\{L\}, \text{with  } L(\boldsymbol{\theta}, \boldsymbol{\phi}) = {R}   + \lambda_{1} \cdot \mathrm{MSE}(\cdot)\\
		+ \lambda_{2} \cdot (1.0 - \mathrm{MS\text{-}SSIM}(\cdot))
		+   \lambda_{3} \cdot \Delta E_{00}^{12}(\cdot),
		\label{eqn:loss}
	\end{aligned}
\end{equation}

where $\lambda_{1}, \lambda_{2}, \lambda_{3}$ are the Lagrangian multipliers for the metrics MSE, MS-SSIM and CIEDE2000 respectively. It should be noted that MSE and MS-SSIM are estimated in the RGB color space. ${R}$ indicates the total bitrate.

\vspace{-8pt}
\subsection{Channel impulse response} \label{ssec:con}
\vspace{-4pt}

In a linear transformation such as the DCT, the reconstructed image can be understood as the superposition of suitably weighted impulse responses or basis functions. But in case of a learned codec, due to the nonlinearity of the synthesis transform, this equality no longer holds, so we can use it as a tool to understand what the network actually learned from the input image. In order to perform such an analysis, we extend the method described in \cite{10018031} to obtain the impulse responses of the color learning model.

Firstly we perform analysis transforms on a given YUV image. In order to obtain an impulse response for each latent channel, we consider a channel index $i$ for luminance, with $i = 0, ...,127$ and $j$ for chrominance, with $j = 0, ...,63$. For each latent channel $i$ or $j$, we choose the single largest value and set the values in all the other channels to zero. These form the “impulses”, with dimensions $1\times1\times128$ for luminance and $1\times1\times64$ for chrominance, which are the inputs to synthesis transforms. The resulting images are the “impulse responses”. Effectively, we have 128 outputs for the luminance branch and 64 for chrominance with dimensions $16\times16$, due to upsampling of the synthesis transforms. Finally they are converted to RGB. For visualization, the impulse responses are arranged in the decreasing order of channel bitrates, measured using prior distribution.

\vspace{-8pt}
\subsection{Implementation details}
\vspace{-4pt}
The model was implemented in Python programming language using the \textit{Tensorflow} framework along with the \textit{Tensorflow compression} \footnote[1]{\url{https://github.com/tensorflow/compression}} library for the entropy models and GDN \cite{balle2016density}. The proposed model has a total of 8,163,051 parameters. Similar to many learned image compression methods, we used standard datasets to train and test our model. The \emph{Coco2017}\cite{lin2014microsoft} test and validation dataset comprising of about 45,000 images was used to train the model for multiple bitrate configurations. As validation data, we used the \emph{CLIC - mobile and professional} training datasets\cite{toderici2020workshop}, containing 586 images. A learning rate of $1e-4$ was used with the Adam\cite{kingma2014adam} optimizer.  The values for Lagrangian multipliers were chosen experimentally as $\lambda_{1}=\{0.001, 0.005, 0.01, 0.02\}$, $\lambda_{2}=\{0.01, 0.12, 2.4, 4.8\}$ and $\lambda_{3}=\{0.024, 0.12, 0.24, 0.48\}$. The training images were randomly cropped into patches of size $256\times256$ and a batch size of 8. As test data, \emph{Kodak}\footnote[2]{\url{https://r0k.us/graphics/kodak/}} (24 images) and \emph{CLIC - professional} validataion dataset (41 images) were used.

\vspace{-8pt}
\section{Results}
\vspace{-4pt}
\label{sec:res}
In this section, we discuss the experimental results and observations that demonstrate the benefits of our work. We compare the performance of our model with other codecs. Additionally, the visualization and analysis of the latent channel impulse responses is also carried out.
\vspace{-8pt}
\subsection{Comparison of R-D curves}
\vspace{-4pt}
In order to compare the compression performance with other codecs, we plot the rate-distortion (R-D) curves for the \emph{Kodak} dataset. The comparison is made with traditional codecs - JPEG and VVC as well as learning based codecs - \texttt{cheng2020} \cite{cheng_learned_2020} and \texttt{bmshj2018}\cite{balle2018variational}. For the learned codecs, pre-trained models from the \textit{CompressAI} library \cite{begaint2020compressai} were used. The R-D curves are shown in Fig. \ref{fig:rd_curves} with PSNR, MS-SSIM and CIEDE2000 for the bitrate range of 0 to 1 bpp. For better readability, we converted MS-SSIM into decibels (dB). We observe that our model outperforms JPEG and is comparable to \texttt{bmshj2018} for all three metrics. In terms of MS-SSIM, the results are close to \texttt{cheng2020}. However, for bitrates greater than 0.5 bpp, it is close to that of VVC. For CIEDE2000, the performance is close to VVC only for the last data point around 0.8 bpp.

\begin{figure}[th]
	\centering
	\begingroup
	\setlength{\tabcolsep}{1pt}
	\begin{tabular}{c c }
		Original & Ours \\
		\includegraphics[width=0.33\linewidth]{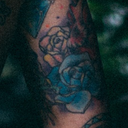} 
		& \hspace{-1pt} \includegraphics[width=0.33\linewidth]{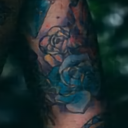} \\
		\small PSNR $\uparrow$, MS-SSIM $\uparrow$, CIEDE2000 $\downarrow$& \small$36.78, 0.98, 2.08$\\
		& \\
		JPEG \cite{125072} & \texttt{cheng2020} \cite{cheng_learned_2020}\\
		\includegraphics[width=0.33\linewidth]{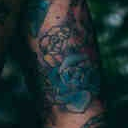} 
		& \includegraphics[width=0.33\linewidth]{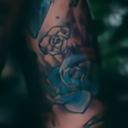} \\
		\small$33.94, 0.95, 2.66$ & \small$36.61, 0.97, 2.22$\\
	\end{tabular}
	\endgroup
	\caption{ Comparison of original image patch and decoded patches from the proposed model, JPEG and \texttt{{cheng2020}}, compressed with a bitrate of around 0.4 bpp.} \label{fig:comp}
\vspace{-16pt}
\end{figure}

\vspace{-8pt}
\subsection{Comparison of reconstruction fidelity}
\vspace{-4pt}
Here we consider a region of interest in the form of a patch from the image \texttt{felix-russell-saw-140699.png}, taken from the \emph{CLIC-professional} validation dataset. We perform a comparison of visual quality with JPEG and \texttt{cheng2020} using configurations that yield a bitrate of around 0.4 bpp. In Fig. \ref{fig:comp}, the original image patch and the decoded image from the proposed model are shown in the first row. In the second row, we have the reconstructed images from JPEG and \texttt{cheng2020}. Block artifcats can be observed in the JPEG image patch. Although the \texttt{cheng2020} image appears perceptually pleasant, on closer inspection, the sharp areas appear smooth and blurred. In terms of color reconstruction, we observe some regions where colors are not preserved and in case of small regions close to a large distribution of a dominant color, there is smudging. For example this is evident when we observe the blue flower in the \texttt{cheng2020} image. However, with our model, we clearly observe better reconstruction quality both in color and structure. The image is also sharper in comparison to the other codecs. This is supported by the objective quality metrics measured for the complete image. 

\begin{figure}[t]
	\centering
	\begin{subfigure}[t]{\columnwidth}
		\centering
		\begin{tabular}[t]{c}
			Luminance \\
			\includegraphics[height=0.39\linewidth]{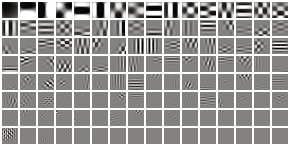} \\
			Chrominance \\
			\includegraphics[height=0.39\linewidth]{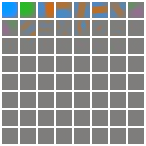}\\
		\end{tabular}
		\caption{Ours: Luminance - 128 channels, Chrominance - 64 channels.}\label{fig:impulse_cl}
		\vspace{4pt}
	\end{subfigure}
	\begin{subfigure}[h]{\columnwidth}
		\centering
		\includegraphics[height=0.39\linewidth]{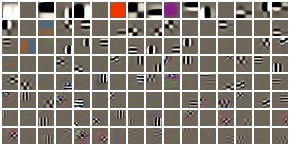}
		\caption{\texttt{cheng2020}  \cite{cheng_learned_2020}: 128 channels.}\label{fig:impulse_cheng}
	\end{subfigure}
	\caption{{Channel impulse responses} for the image \texttt{ClassA\_8bit\_Bike\_2048x2560\_8b\_RGB.png}. The sub-images of size $16\times16$ are arranged in a grid based on the decreasing order of channel bitrate contribution.}
	\label{fig:impulse}
	\vspace{-16pt}
\end{figure}

\vspace{-10pt}
\subsection{Comparison of channel impulse responses}
\vspace{-4pt}

Based on the process explained in Sec. \ref{ssec:con}, we compute the channel impulse response of the proposed model and compare the results to that of \texttt{cheng2020}. In our experiment, we use an image from JPEG XL test data - \texttt{ClassA\_8bit\_Bike\_2048x2560\_8b\_RGB.png}. The results are visualized in Fig.\ref{fig:impulse}.

From the impulse responses of \texttt{cheng2020} shown in Fig. \ref{fig:impulse_cheng}, we observe a combination of patterns for structure and color. In some channels, we see abstract patterns that contain both. The impulse responses for the color learning model are shown in Fig. \ref{fig:impulse_cl}. The upper image represents luminance images and the lower one indicates chrominance. We can clearly see the separation of structure in the luminance images and color in the chrominance images in comaprison to \texttt{cheng2020}. It is also quite interesting to see the similarity of some channels in \texttt{cheng2020} and numerous luminance impulse responses of our model to the basis functions of an orthogonal transform, such as DCT. Thus, we can confirm the findings in \cite{10018031}, that the transforms of learned image codecs can be interpreted as the non-linear counterparts of orthogonal transforms. However, we plan to develop a method to quantitatively measure the similarity between the transforms of a learned image codec and a linear transform coder in a follow-up work.
\vspace{-14pt}
\section{Conclusion} \label{sec:con}
\vspace{-6pt}
In this work, we develop a deep neural network for the task of image compression with the idea of capturing structural and color information separately. We optimize the model using RDO with an additional metric - CIEDE2000 to measure color difference. The comparison with other codecs demonstrates the structural and color fidelity of the proposed model. The channel impulse response experiment validates that our model captures structural and color information separately. This can be used to perform cross-component prediction and enable residual coding, that could also be extended to learned video compression. The presented method can be easily adopted by state-of-art methods to enhance color fidelity.

\balance
\bibliographystyle{IEEEbib}
\bibliography{refs}

\end{document}